\documentstyle[12pt]{article}

\newcommand\beq{\begin{equation}}
\newcommand\eeq{\end{equation}}
\def\beqa{\begin{eqnarray}}
\def\eeqa{\end{eqnarray}}
\def\bega{\begin{array}}
\def\enda{\end{array}}
\def\non{{\nonumber}}
\def\p{^\prime}

\begin{document}

\title{RENORMALIZATION AMBIGUITIES IN CASIMIR ENERGY}
\author{Luiz C. de Albuquerque\thanks{e-mail:{\it claudio@fma1.if.usp.br}}\\
\\{\small\it Instituto de F\'\i sica, Universidade de S\~ao\ Paulo}\\
{\small\it P.O. Box 66318, 05389-970 S\~ao\ Paulo, SP,  Brazil}}
\date{August 1997}
\maketitle
\vspace{-13cm}
\vspace{13cm}
\begin{abstract}

{\sl Some questions were recently raised about the equivalence of two 
methods commonly used to compute the Casimir energy:
the mode summation approach and the one-loop effective potential. 
In this respect, we argue that the scale dependence 
induced by renormalization effects, displayed by the effective potential 
approach, also appears in the MS method.}

\end{abstract}
\vspace{3cm}

\bigskip

\newpage

\baselineskip=20 pt

Casimir \cite{Casimir}
showed that two neutral perfectly conducting parallel plates placed in 
the vacuum  attract each other, due to zero-point oscillations
of the electromagnetic field (field strength fluctuations).  His starting
point resembles an old idea of Euler and Heisenberg \cite{Euler}. They
used the zero-point oscillations of the Dirac field 
(charge fluctuations) in an external  field to define 
an effective action to the electromagnetic field. The common
feature is the summation of the energies associated with the  
vacuum fluctuations of the constrained quantum field
to define the vacuum energy and the effective action.
The  effective action was later reinterpreted in terms
of the generating  functional of 1PI Green's functions.
The effective action at one-loop order is formally equivalent to the
older definition of Euler and Heisenberg \cite{Weinberg}, and 
the vacuum energy expression obtained from it is also formally
identical to the mode summation method used by Casimir. 

The formal expression of the vacuum energy 
is ultraviolet (UV) divergent. The existence of infinite
contributions in renormalizable field theories can be handled
by the  regularization/renormalization process. This
procedure introduces two possible  sources of ambiguities.
The separation in a finite part plus an infinite
contribution is  arbitrary since many different subtractions  
are possible. The regularization procedure also introduces a
dimensionfull parameter, the renormalization scale. The renormalization
condition  fix the finite part of the counterterms, and 
it may or it may not discard the
renormalization scale.  We  deserve the name Casimir
energy for the regularized vacuum energy satisfying some definite
renormalization condition. The requirement of a zero
vacuum energy for flat unconstrained Minkowski space-time
is compatible with normal ordering concepts \cite{Kay}.
If this condition succeed in discard the
infinite contributions and the renormalization scale,
the Casimir energy is uniquely defined. This happens in the
electromagnetic Casimir effect.\linebreak
Otherwise the Casimir energy is
ambiguous and, since the pressure on the plates is the
unique measurable quantity, the result is non-predictive 
\cite{Blasi}. This may occurs for instance as a consequence
of a mass term \cite{Luiz}, or deviations from the plane geometry
\cite{Blasi,Deutsch,Blau}. If the interest is not on the computation
of Casimir forces, but in the contribution of the Casimir effect 
for other process, like the bag pressure in  MIT Bag model
calculations \cite{Blau},  this ambiguity signals the existence of a 
new, phenomenological, free parameter \cite{Blasi}.
This observation is of importance in studies of liquid Helium films
\cite{Schw}.

It was recently argued that
the effective potential method is more reliable
than the mode summation (MS) approach in the computation of 
the Casimir energy \cite{Myers}: in some cases
the MS approach may neglect a renormalization scale dependent
contribution. The purpose of this paper is to show 
that the same scale dependence
is expect in both methods (one-loop effective potential and MS).
This indeed was shown in \cite{Blau} for a massless scalar field.
Here, a global analysis \cite{Actor} is made of the massive 
scalar case  for flat parallel plates.  
As a by-product, we also establish the analytical equivalence
between  regularization prescriptions of the MS formula and 
regulated definitions of the effective potential.
We  do not claim that the Casimir energy 
is unambiguous defined in all situations: 
since  subtractions are necessary, the Casimir
energy will depend on the subtraction employed. 

Let $\phi(x)$ be a real massive scalar field, 
with a $\lambda\phi^4/4!$ self-coupling\footnote{\footnotesize
$D$ dimensional Euclidean space with positive signature.
$x^{\mu}=(t,x_1,\vec{x}_\perp), \bar{x}^\mu=(t,\vec{x}_\perp),
\bar{k}^\mu=(k_0,\vec{k}_\perp)$.}.
The imposition of Dirichlet boundary 
conditions (DBC) in $x_1$, i.e. 
$\phi(\bar{x},x_1=0)=0=\phi(\bar{x},x_1=L)$,
leads to the discrete set 
$k_1={\pi n\over L}$ ($n\in {\cal N}$), plus plane waves in the 
other $d=D-1$ dimensions. The DBC simulate the perfect conducting
plates in the Casimir apparatus.

The one-loop effective potential is given by the sum of the
1PI one-loop diagrams with zero external momenta.
After summation \cite{Cole}, we have in the unconstrained space

\beq
\label{e1}
V^{(1)}(\bar{\phi})=\frac{1}{2}\int \frac{d^D k}{(2\pi)^D}
\Biggl[\ln \bigl(k^2+M^2(\bar{\phi})\bigr)-
\ln\bigl(k^2+m^2\bigr)\Biggr]+\delta V^{(1)}(\bar{\phi}),
\eeq
where $M^2= m^2+{\lambda\over2}\bar{\phi}^2$, 
$\bar{\phi}$ is a constant background field, and $\delta
V^{(1)}(\bar{\phi})$ is the counterterm of order $\hbar$.
The second term in the square brackets comes from the normalization
of the generating functional. The minimum of the effective potential
gives the vacuum energy density \cite{Cole,Zuber}.  We study the case without
spontaneous symmetry breaking (SSB), i.e. $m^2>0$ (the
case with SSB cannot be discussed in a global framework
\cite{Toms}).  Thus (the normalization factor and the counterterm 
are not written)
 
\beq
\label{0}
{\cal E}_0 (L)=V^{(1)}(\bar{\phi}=0)=
\frac{1}{2}\hbox{$\sum$}\!\!\!\!\!\!\!\int^{(L)}_{(D-1)}\,
\ln \Bigl(\bar{k}^2 +\bigl(\frac{\pi n}{L}\bigr)^2 +m^2\Bigr),
\eeq
where the DBC was imposed, and our notation is

\beq
\label{01}
\hbox{$\sum$}\!\!\!\!\!\!\!\int^{(L)}_{(D-1)}\,\equiv{1\over L}
\sum_{n=1}^{\infty}\int _{-\infty}^{\infty}
{d^{D-1}\bar{k}\over (2\pi)^{D-1}}=\,{\rm Tr}_k^{(D)}\mid_{\rm DBC}.
\eeq
${\rm Tr}_k^{(D)}$ stands for the trace in the $D$ dimensional
momentum ($k$-)space. Integration over $k_0$ gives the MS expression

\beq
\label{e2}
{\cal E}_0(L)=\frac{1}{2}\hbox{$\sum$}\!\!\!\!\!\!\!\int^{(L)}_{(D-2)}\,
\Bigl(\vec{k}_{\perp}^2+ \bigl(\frac{\pi n}{L}\bigr)^2+
m^2\Bigr)^{\frac{1}{2}}.
\eeq
${\cal E}_0$ is the sum of all possible zero-point 
oscillations centered about $\bar{\phi}=0$.

The equivalence showed above involves manipulations with
ill-defined quantities. It is therefore necessary to give a 
mathematical definition to the righ-hand side (RHS) of
equations (1) and (4). 
To start, the one-loop effective potential is written 
as\footnote{\footnotesize  Note that 
$\ln\det\,{\cal A}_{x,y}={\rm Tr}^{(D)}\ln\,{\cal A}_{x,y}
=\Omega_D {\rm Tr}_k^{(D)} \ln\,{\cal A}_k=\Omega_D\ln \det_k\, 
{\cal A}_k$, where $\Omega_D$ is a $D$-dimensional normalization 
volume, and ${\cal A}_{x,y}=(-\Box+m^2)\delta^{D}(x-y)$.}

\beq
\label{3}
V^{(1)}(\bar{\phi}=0)={1\over2}\ln\;{\det}_{k}
\Biggl({{\cal O}\over \mu^2}\Biggr)_{\rm DBC},
\eeq
where ${\cal O}(\bar{\phi}) =k^2+m^2$,  and $\mu$ is a
mass scale used to normalize the determinant.  Now a regulated version
of eq.(5) is supplied by the generalized $\zeta$ function technique
\cite{Hawking},

\beqa
\label{4}
{\det}_{k} \Biggl({{\cal O}\over \mu^2}\Biggr)_{\rm DBC}
&=&\exp \;\Biggl[\;
-{\partial\over \partial s}\zeta_{{{\cal O}\over \mu^2}}(s)\mid_{s=0}
\;\Biggr]\\
\label{4b}
& &=\Bigl(\mu^2\Bigr)^{-\zeta_{{\cal O}}(0)}
\exp \;\Biggl[\;
-{\partial\over \partial s}\zeta_{{\cal O}}(s)\mid_{s=0}
\;\Biggr].
\eeqa
where 

\beq
\label{5}
\zeta_{{{\cal O}\over \mu^2}}(s)=
{\rm Tr}_k^{(D)}\;\Bigl({{\cal O}\over \mu^2}\Bigr)^{-s}\mid_{\rm DBC}\,
\equiv \hbox{$\sum$}\!\!\!\!\!\!\!\int^{(L)}_{(D-1)} \;\;
\Bigl(\Lambda_n\mu^{-2}\Bigr)^{-s}
\eeq
is the generalized $\zeta$ function associated with the $D$-dimensional
operator $\frac{{\cal O}}{\mu^2}$, subjected to the DBC, with
eingenvalues given by $\mu^{-2}\Lambda_n=\bar{k}^2 +\bigl({\pi n\over L}\bigr)^2+m^2 $. $\zeta_{{\cal O}}(s)$ is the $\zeta$ function
of the operator ${\cal O}$, with the scale $\mu$ factored out.
Using equations (\ref{0}), (\ref{4}), and (\ref{4b}), 
we obtain the two equivalent forms\footnote{\footnotesize
The simplified notation is: 
${\partial\over \partial s}\zeta_{{{\cal O}\over
\mu^2}}(s)\mid_{s=0}\equiv \zeta\p_{D}(0;\mu)$.}

\beq
\label{100}
{\cal E}_0 (L)=-{1\over2}{\partial\over \partial s}\zeta_D(s;\mu)\mid_{s=0}
=-{1\over2}\Biggl[\zeta\p_D(0)+ \zeta_D(0)\ln\mu^2\Biggr].
\eeq

An analytical continuation to the whole complex $s-$plane must be 
done before we take the limit $s\rightarrow0$.
It can be shown that for an elliptic, positive second-order operator, $\zeta_D(s)$ is a meromorphic function with only simple poles,
in particular analytic at $s=0$ \cite{Hawking}. Hence, ${\cal E}_0(L)$
defined in (\ref{100}) is a finite quantity.

The  scale contribution to the 
Casimir energy comes from
the $\zeta (0)$ term in eq. (9). It is the absence of an analogous
term in the naive MS formula, eq.(\ref{e2}), that leads
Myers \cite{Myers} to introduce the notion of \lq\lq zero-point
anomaly''. We will show below that the same scale behavior 
appears also in the MS approach. The key point is: 
by virtue of the formal relation derived above, 
a proper definition (i.e. regularization) of the effective potential
leads to a definition of the MS formula eq.(4).
To proceed along this way, we  integrate over 
$k_0$ in eq. (\ref{5}). This gives

\beq
\label{7}
\zeta_D(s;\mu)=-\frac{\mu f(s)}{\Gamma(s)}\,
\hbox{$\sum$}\!\!\!\!\!\!\!\int^{(L)}_{(D-2)}\,
\Biggl[{{\vec{k}_{\perp}^2 +\bigl({\pi n\over L}\bigr)^2
+m^2}\over \mu^2}
\Biggr]^{\frac{1}{2}-s},
\eeq  

\noindent $f(s)= -\Gamma\biggl(s-\frac{1}{2}\biggr)/(2\sqrt{\pi})$ is an
irrelevant factor ($f(0)=1$). Now, using (\ref{100}) we obtain for 
${\cal E}_0(L)$ (plus counterterms):

\beqa
\label{8}
{\cal E}_0(L)&=&{1\over2}\mu\,{\partial\over \partial s}\;
\Biggl\{{1\over \Gamma(s)}\hbox{$\sum$}\!\!\!\!\!\!\!
\int^{(L)}_{(D-2)}\;\;
\Bigl(\lambda_n\mu^{-2}\Bigr)^{\frac{1}{2}-s}\Biggr\}_{s=0}\non\\
& =&{1\over2}\mu\,{\partial\over \partial s}\Biggl\{{1\over \Gamma(s)}
\zeta_{D-1}\Bigl(s-\frac{1}{2};\mu\Bigr)\Biggr\}_{s=0},
\eeqa
where $\lambda_n= \vec{k}_{\perp}^2 +\bigl({\pi n\over L}\bigr)^2
+m^2 \equiv \omega_n^2$ are the eigenvalues of the
$(D-1)$-dimensional (reduced) operator $\tilde{\cal O}= \vec{k}+m^2$.
We call $\zeta_{D-1}(s)$ the \lq\lq reduced" $\zeta$ function.

For a function $G(s)$ analytic at $s=0$, we use the approximation
$1/\Gamma(s)\approx s+\gamma s^2+O(s^3)$ to deduce 
${\partial\over \partial s}\;{G(s)\over \Gamma(s)}\vert_{s=0}=G(0)$.
Suppose that $\zeta_{D-1}\bigl(s-\frac{1}{2};\mu\bigr)$ is analytic
at $s=0$; Then, we will obtain 

\beq
\label{13}
{\cal E}_0 (L)={1\over2}\mu\;\zeta_{D-1}\Bigl(-\frac{1}{2};\mu\Bigr)
={1\over2}\hbox{$\sum$}\!\!\!\!\!\!\!\int^{(L)}_{(D-2)} \;\;
\omega_n.
\eeq
This is exactly the non-regulated MS formula for the vacuum energy,
eq.(\ref{e2}), in the present notation. 
It is clear that eq.(\ref{8}) is  another way to write
eq.(\ref{100}). However, formula (\ref{8})  by itself is also a 
regularized expression for the vacuum energy in a MS like form
(perhaps a strange one!). The UV divergences, associated with the 
high-frequency modes in (\ref{13}), appear as poles in
$\zeta_{D-1}\bigl(s-\frac{1}{2};\mu\bigr)$ when we take $s\rightarrow0$.
In general $\zeta_{D-1}\bigl(s-\frac{1}{2};\mu\bigr)$ is non
regular at $s=0$\footnote{\footnotesize Note that the entire expression
on the RHS of (\ref{8}) is analytic at $s=0$, as well as
$\zeta_{D-1}(s;\mu)$}.  Thus, we can
interpret eq.(11) as a MS regularized formula for the vacuum energy.

A Mellin transform can be made to relate
the generalized $\zeta$ function to the trace of the heat kernel,
$Y(t)\equiv {\rm tr}\, e^{-t\tilde{\cal O}\mu^{-2}}$.
As is known \cite{DeWitt}
the heat kernel possess an asymptotic expansion for small $t$.
Using this expansion,  we get the pole structure
of $\zeta_{D-1}(s;\mu)$ \cite{Birrel}

\beq
\label{16}
\zeta_{D-1}(s;\mu)={1\over (4\pi)^{\frac{D-1}{2}}\Gamma(s)}
\Biggl\{\sum_0^\infty{C_j\over {s-(\frac{D-1}{2}-j)}}+F(s)\Biggr\}.
\eeq

\noindent $F(s)$ is an entire analytic function of $s$. We see that
$\zeta_{D-1}(s;\mu)$ is a meromorphic function of $s$, with
simple poles at $s=\frac{1}{2},1,\frac{3}{2},...\frac{D}{2}$
with residua given by the coefficients $C_j$. 

Using the asymptotic expansion (13) in eq. (\ref{8}), we obtain 

\beqa
\label{17}
{\cal E}_0(L)&=&{\mu\over2 (4\pi)^{\frac{D}{2}}}{\partial\over \partial s}
{1\over \Gamma(s)}\Biggl\{
\sum_{0 \atop j\neq D/2}^\infty {C_j\over {s-(\frac{D}{2}-j)}}+
F(s-\frac{1}{2})
+ {C_{\frac{D}{2}}\over s} \Biggr\}\non\\
& &=\bar{\cal E}_0(L) - {\psi(1)\over 2(4\pi)^{\frac{D}{2}}}\,
\mu C_{\frac{D}{2}}(\mu).
\eeqa
where $\bar{\cal E}_0(L)$ comes from the regular part inside the
curly brackets in the first line. The combination
$\mu C_{\frac{D}{2}}(\mu)$ do not depend on $\mu$ (Appendix A
relates the various $\zeta$ functions appearing in the text).

To see the effect of a change of scale, $\mu\rightarrow\mu\p$,
eq.(\ref{5}) can be used to obtain

\beq
\label{20}
\zeta_D(s;\mu\p)=\Biggl({\mu\p\over \mu}\Biggr)^{2s}
\zeta_D(s;\mu).
\eeq
Using the equations (\ref{8}) and (\ref{20}), we have
\beq
\label{21}
{\cal E}_0(\mu\p)=
{\cal E}_0(\mu)-{\mu\over (4\pi)^{\frac{D}{2}}}C_{\frac{D}{2}}(\mu)\,
\ln\Bigl({\mu\p\over \mu}\Bigr).
\eeq

The scale dependence is logarithmic and proportional to the  
coefficient  $C_{\frac{D}{2}}$. This is already clear 
from eq.(\ref{100}) since $\zeta_D(0)\propto \mu C_{\frac{D}{2}}(\mu)$.
It is obvious from our approach that the
same scale behavior is found both in the effective potential
method (with the $\zeta$ function regularization) and
in the MS approach (with the regularization prescription (\ref{8})).
From another point of view, the $\zeta$ function definition of
the effective potential is equivalent to 
the analytical regularization (11) of the MS formula.

For conformally invariant theories in flat space-time and  
flat parallel plates, $C_{\frac{D}{2}}=0$ and  $\mu$ disappears 
identically \cite{Blau}. In the $\zeta$ function approach, 
one can define the Casimir energy by demanding that the vacuum energy 
disappears in the limit $L\rightarrow\infty$, thus fixing the finite
part of the constant counterterm in (\ref{e1}) \cite{Toms}.
Another conventional choice, related to the previous one,
is to define the  Casimir  energy (density) as  \cite{Phys.Rep.}

\beq
\label{18}
{\cal E}_c={\cal E}_0(L)-{\cal E}_0( L\rightarrow\infty).
\eeq
This prescription
is automatically achieved if we use the normalization factor
in (\ref{e1}), calculated at $L\rightarrow\infty$.
The renormalization condition (\ref{18}) is 
physically reasonable since it leads to a zero vacuum energy for
flat unconstrained space-time, consistent with the usual
normal ordering  prescription.

In the general case 
we expect a dependence on $\mu$ in the Casimir energy.
In the simple model discussed here
the prescription (\ref{18}) (or the equivalent one) can not
fix the scale for odd dimensional  space-times:

\beqa
\label{18B}
& &\mu C_{\frac{D}{2}}^{(+)}(\mu)=(4\pi)^{\frac{D}{2}}\zeta^{(+)}_D(0)=
\frac{(-1)^{\frac{D}{2}}}{\Bigl(\frac{D}{2}\Bigr)!}m^D,\non\\
& &\mu C_{\frac{D}{2}}^{(-)}(\mu)=(4\pi)^{\frac{D}{2}}\zeta^{(-)}_D(0)=
-\frac{(-1)^{\frac{D-1}{2}}}{\Bigl(\frac{D-1}{2}\Bigr)!}
\frac{\sqrt{\pi}m^{D-1}}{L},
\eeqa
where $+$ is for even and $-$ is for odd $D$-dimensional space-time,
see eq. (A.8).
The renomalization condition (\ref{18}) fails to give an 
unambiguous result for odd $D$ since
$C_{\frac{D}{2}}^{(-)}\rightarrow0$ as $L\rightarrow\infty$.
$\zeta(0)$ is closely related to the trace anomaly \cite{Birrel}.
For unconstrained  fields or periodic boundary conditions (PBC), 
$\zeta_D^{(-)}(0)=0$, even for $m\neq0$ (indeed,
$\zeta_D^{(-)}(0)=0$ for $L\rightarrow\infty$). Hence,
result (18) seems to indicate a breakdown of 
scale invariance for odd $D$ induced by the DBC.
This point deserves a further study, which however goes beyond the 
present purpose. The value of $\zeta^{(+)}_D(0)$
is the same as in the unconstrained space case
($L\rightarrow\infty$). This is also true in
the PBC case for any space-time
dimension (including the particular cases of a compactified space dimension, 
or finite temperature
field theory in the Euclidean time). This fact is generalized in the
statement that PBC do not introduce new ultraviolet 
structures  in field theory, besides the usual ones. As eq.(18)
indicates this is no longer true in the DBC case \cite{Luiz}.

Thus, to define a finite  Casimir energy we have to impose a 
renormalization condition (as in eq.(\ref{18})), besides a
regularization method. Of course, different renormalization conditions
may lead to different (finite) Casimir energies.
However, the logarithmic dependence displayed in eq.(\ref{21})
is a general  feature, within the class of analytical regulators
discussed here\footnote{\footnotesize The same is true in the 
dimensional regularization method.}.
Many regularization prescriptions for the MS
formula (\ref{e2}) are possible, and indeed are 
fairly used \cite{Phys.Rep.}. 
For instance, starting from the unregulated MS formula (\ref{13}), 
Blau et al.  \cite{Blau} used the following analytical regularization   
(to be compared with eq. (11))

\beq
\label{22}
{\cal E}_0(s)={1\over2}\mu
\hbox{$\sum$}\!\!\!\!\!\!\!\int^{(L)}_{(D-2)} \;\;
\Bigl(\lambda_n\mu^{-2}\Bigr)^{\frac{1}{2}-s}
={1\over2}\mu\,\zeta_d \Bigl(s-\frac{1}{2};\mu\Bigr).
\eeq
In this case, the pole in $s=0$ do not cancel against $\Gamma(s)$,
and the coefficient $C_{\frac{D}{2}}$ turns into an obstacle to give a
finite Casimir energy. The total
energy is  finite, because of its effects over the gravitational 
field. Hence, the bare action  must contain a term proportional to
$C_{\frac{D}{2}}$. In the minimal subtraction scheme
proposed in \cite{Blau} the pole is simply removed, and the 
Casimir energy is  defined by (P=principal value)

\beq
\label{23}
{\cal E}_c\equiv \lim_{s\rightarrow0}{1\over2}\Bigl\{{\cal E}_0(+s)+
{\cal E}_0(-s)\Bigr\}
\equiv {\mu\over2}\, {\rm P}\,\zeta_d\Bigl(s-\frac{1}{2};\mu\Bigr).
\eeq
Although they used a completely different  regularized MS
expression  (compare eq.(\ref{22}) with eq.(\ref{8})) and more important, 
another renormalization condition (eq.(\ref{23}) instead of eq.(\ref{18})), 
Blau et al. \cite{Blau} derived  the formula (\ref{21}) 
for the  scale dependence of the Casimir energy in the massless case.
Clearly, the Casimir energy computed according equations (19) and (20) is
not necessarily identical to the one computed using 
equations (11) and (17).
For instance, in the model studied here a direct application
of equations (19), (20) and (A.7)leads to

\beq
\label{0001}
{\cal E}_c=-\frac{1}{(4\pi)^{\frac{D}{2}}}\Biggl[\,
g_{\pm}^\mu +2\Biggl(\frac{m}{L}\biggr)^{\frac{D}{2}}
\sum_{n=1}^\infty n^{-\frac{D}{2}} K_{\frac{D}{2}}\bigl(2mL n\bigr)
\Biggr],
\eeq
where                          

\begin{eqnarray}
\label{0002}
& &g_{+}^\mu=-\frac{\sqrt{\pi}m^{D-1}}{2L}
\Gamma\Bigl(-\frac{D-1}{2}\Bigr)
+\frac{(-1)^{\frac{D}{2}}}{\bigl(\frac{D}{2}\bigl)!}m^D
\ln\,\frac{\mu}{m},\\
& &g_{-}^\mu=\frac{m^{D}}{2}\Gamma\Bigl(-\frac{D}{2}\Bigr)
+\frac{(-1)^{\frac{D-1}{2}}}{\bigl(\frac{D-1}{2}\bigr)!}
\frac{\sqrt{\pi}m^{D-1}}{2L}
\ln \,\frac{\mu}{m}.\nonumber
\end{eqnarray}
This result is to be contrasted with that  obtained by an application
of equations (11) and (17) \cite{Luiz}

\beq
\label{0003}
{\cal E}_c=-\frac{1}{(4\pi)^{\frac{D}{2}}}\Biggl[\,
\tilde{g}_{-}^\mu +2\Biggl(\frac{m}{L}\biggr)^{\frac{D}{2}}
\sum_{n=1}^\infty n^{-\frac{D}{2}} K_{\frac{D}{2}}\bigl(2mL n\bigr)
\Biggr],
\eeq
where $\tilde{g}_{-}^\mu=g_{-}^\mu-\frac{m^D}{2}
\Gamma\bigl(-\frac{D}{2}\bigr)$.

Finally, we will show that the MS regularization proposed in
\cite{Blau} has an analytical counterpart in the effective potential 
method. Using the formula

\beq
\label{002}
\ln\Bigl(\frac{b}{a}\Bigr)=\int_0^\infty \frac{dx}{x}
\bigl[e^{-ax}-e^{-bx}\bigr],
\eeq
we may recast $V^{(1)}(\bar{\phi})$ in eq. (1) as

\beq
\label{003}
V^{(1)}(\bar{\phi})=-\frac{1}{2}\int_0^\infty
\frac{d\tau}{\tau} {\rm Tr}_k^{(D)}\Biggl[e^{-\tau(k^2+M^2)}
-e^{-\tau(k^2+m^2)}\Biggr].
\eeq
This is essentially the Schwinger formula (SF) applied to the Euclidean
effective potential \cite{Zuber}.
We proved in \cite{LC1} the equivalence between the SF and the MS
formula. We will repeat only a few steps
of the argument of \cite{LC1}, but now introducing a mass scale $\mu$.

The regularized version of the SF that will be used is \cite{Farina}

\beq
\label{24}
V^{(1)}(\bar{\phi}=0)\mid_{\rm DBC}
=-{1\over2}\int_0^\infty d\tau\, \tau^{s-1} 
\hbox{$\sum$}\!\!\!\!\!\!\!\int^{(L)}_{(D-1)} \;
e^{-\tau(\bar{k}^2+\frac{\pi^2 n^2}{L^2}+m^2)\mu^{-2}},
\eeq
where $s$ is large enough to make the integral well defined, and
$\mu$ is  a mass scale.
In this approach, we first compute the integral, then make 
an analytical continuation to the
whole complex $s-$plane, and finally the limit $s\rightarrow0$
is carefully taken. From eq. (\ref{24}) we obtain

\beq
\label{25}
{\partial V^{(1)}\over \partial m^2}=
{1\over 2\mu^2} 
\hbox{$\sum$}\!\!\!\!\!\!\!\int^{(L)}_{(D-1)} \;
\int_0^{\infty}d\tau \tau^{s}e^{-\tau(\bar{k}^2 +{n^2\pi^2\over L^2}
  +m^2)\mu^{-2}},
\eeq
Using the definition of the Euler Gamma function, 
the integration over $\tau$ is readily done. 
Then, we integrate over $k_0$ to obtain

\beq
\label{26}
{\partial V^{(1)}\over \partial m^2}=
\frac{\Gamma({1\over2})}{4\pi\mu}
\,\hbox{$\sum$}\!\!\!\!\!\!\!\int^{(L)}_{(D-2)}\,
\Biggl[{{\vec{k}_{\perp}^2 +\bigl({\pi n\over L}\bigr)^2
+m^2}\over \mu^2}\Biggr]^{-(\frac{1}{2}+s)}.
\eeq 
Integrating on $m^2$ and identifying 
${\cal E}_0=V^{(1)}(\bar{\phi}=0)$,
we finally obtain (apart from an irrelevant additive constant)
   
\beq
\label{27}
{\cal E}_0(s)={1\over2} g(s)\mu\,
\hbox{$\sum$}\!\!\!\!\!\!\!\int^{(L)}_{(D-2)} \;
\Biggl[{{\vec{k}_{\perp}^2 +\bigl({\pi n\over L}\bigr)^2
+m^2}\over \mu^2}
\Biggr]^{\frac{1}{2}-s},
\eeq 
where  $g(s)=\frac{1}{1-2s}\frac{\Gamma(s+{1\over2})}{\sqrt{\pi}}$. Taking $g(0)=1$, we have

\beq
\label{28}
{\cal E}_0(s)={1\over2}\mu\,
\hbox{$\sum$}\!\!\!\!\!\!\!\int^{(L)}_{(D-2)} \;\;
\Bigl(\lambda_n\mu^{-2}\Bigr)^{\frac{1}{2}-s}
={1\over2}\mu\,\zeta_d\Bigl(s-\frac{1}{2};\mu\Bigr),
\eeq
which is just eq.(\ref{22}). Hence, the analytical regularization
used in the modified SF, eq.(\ref{24}), is  the equivalent
in the effective potential method of the regularized
MS formula of Blau et al. \cite{Blau}. 
We can summarize: To each definition of the effective potential
corresponds a regularized definition of the MS formula.

\bigskip

\renewcommand{\theequation}{A.\arabic{equation}}
\setcounter{equation}{0}

\begin{center}
{\large\bf Appendix A}
\end{center}

The generalized $\zeta$ function $\zeta_{{{\cal O}\over \mu^2}}(s)
=\zeta_D(s;\mu)$ associated with the $D$-dimensional
operator ${\cal O}$ is related to the \lq\lq reduced''
$\zeta$ function $\zeta_{D-1}(s;\mu)$ by 

\beq
\label{B1}
\zeta_D(s;\mu)=\mu f(s)\frac{1}{\Gamma(s)}
\zeta_{D-1}(s-\frac{1}{2};\mu),
\eeq
where $f(s)$ is defined below eq. (\ref{7}). 
The $\zeta$ function $\zeta_D(s;\mu)$ possess an  expansion
of the same sort of the \lq\lq reduced'' $\zeta_d(s;\mu)$,
see eq. (\ref{16}). Let $\tilde{C}_j(\mu)$ be the coefficients
of the expansion of $\zeta_D(s;\mu)$; a straightforward
calculation gives

\beqa
\label{B2}
& &\zeta_D(0;\mu)=\frac{1}{(4\pi)^{\frac{D}{2}}}
\tilde{C}_{\frac{D}{2}}(\mu),\non\\
& &\tilde{C}_{\frac{D}{2}}(\mu)=\mu C_{\frac{D}{2}}(\mu).
\eeqa

$\zeta_D(s)$ is the $\zeta$ function of the operator ${\cal O}$.
It is  possible to relate  $\zeta_D(s)$ and $\zeta_D(s;\mu)$.
If $\bar{C}_j$ are the coefficients of the expansion of
$\zeta_D(s)$, it can be  show that
$\tilde{C}_{\frac{D}{2}}(\mu)=\bar{C}_{\frac{D}{2}}$.
Thus, by eq. (\ref{B2}) the combination
$\mu C_{\frac{D}{2}}(\mu)$ do not depend on $\mu$:

\beq
\label{B3}
\mu C_{\frac{D}{2}}(\mu)=(4\pi)^{\frac{D}{2}}\zeta_D(0).
\eeq

To compute  $\zeta_D(0)$,  we integrate eq. (8) (without the scale
$\mu$) to obtain

\begin{equation}
\label{B4}
\zeta_D(s)=\frac{1}{(4\pi)^{\frac{D-1}{2}}}\frac{\Gamma(s-\frac{D-1}{2})}
{\Gamma(s)}\frac{1}{L} E_1^{m^2}(s-\frac{D-1}{2};\frac{\pi^2}{L^2}),
\end{equation}
where  we introduced the modified inhomogeneous Epstein 
function 

\begin{equation}
\label{B5}
E_{1}^{c^2}(s; a)=\sum_{n=1}^\infty
\Bigl( a n^2+ c^2\Bigr)^{-s},
\end{equation}
with $a,c^2\,>\,0$. The summation above  converges for 
${\cal R}\,s>\frac{1}{2}$.

The analytical continuation to the whole complex $s-$plane
of $E_1^{c^2}(s;a)$ is given by \cite{Ambjorn}

\begin{eqnarray}
\label{B6}
E_1^{c^2}(\nu,a^2)&=&-\frac{1}{2}c^{-2\nu}+\frac{1}{2}
\frac{\sqrt{\pi}}{a}\frac{\Gamma(\nu-\frac{1}{2})}
{\Gamma(\nu)}c^{1-2\nu}\non\\
& &\quad +2\frac{\sqrt{\pi}}{a}\Bigl(\frac{c a}
{\pi}\Bigr)^{1/2-\nu}\frac{1}{\Gamma(\nu)}\,
\sum_{n=1}^{\infty}n^{\nu-\frac{1}{2}}\,K_{\frac{1}{2}-s}\Bigl(
\frac{2\pi c}{a}n\Bigr).
\end{eqnarray}

Using  (\ref{B6}) in (A.4), we obtain
\begin{eqnarray}
\label{B7}
\zeta(s)&=&\frac{1}{(4\pi)^{\frac{D-1}{2}}L}\frac{1}{\Gamma(s)}
\Biggl[\,
-\frac{m^{D-1-2s}}{2}
\Gamma\Bigl(s-\frac{D-1}{2}\Bigr) +
\frac{L m^{D-2s}}{2\sqrt{\pi}}
\Gamma\Bigl(s-\frac{D}{2}\Bigr)\non\\
& &\qquad+\frac{2L}{\sqrt{\pi}} \biggl(\frac{m}{L}\biggr)^{\frac{D}{2}-s}
\sum_{n=1}^{\infty}n^{s-\frac{D}{2}}\,K_{\frac{D}{2}-s}
\bigl(2mLn\bigr)\Biggr].
\end{eqnarray}

The pole structure of $\zeta(s)$ is given by 
$\Gamma(s-\frac{D}{2})$ (simple pole for even $D$ 
and $s\rightarrow0$), and $\Gamma(s-\frac{D-1}{2})$ ( odd $D$
and $s\rightarrow0$). Since $\Gamma(s\rightarrow0)
\rightarrow\infty$, $\zeta(0)$ comes from the residue 
of the poles. Using $z\Gamma(z)=\Gamma(z+1)$, we obtain
($+$ corresponds to $D$ even, and $-$ to $D$ odd)

\begin{equation}
\label{B8}
\zeta^{(+)}_D(0)=\frac{(-1)^{\frac{D}{2}}}{(4\pi)^{\frac{D}{2}}
(\frac{D}{2})!}
m^D,\;\;\;\;
\zeta_D^{(-)}(0)=-\frac{1}{2}\frac{(-1)^{\frac{D-1}{2}}}
{(4\pi)^{\frac{D-1}{2}}(\frac{D-1}{2})!}
\frac{m^{D-1}}{L}
\end{equation}
Apart from a minus sign and a factor of 2, $\zeta_D^{(-)}(0)$
resembles $\zeta_D^{(+)}(0)$ for $D-1$ space; the extra factor of $L$ 
in $\zeta_D^{(-)}(0)$ is  seem to be necessary for dimensional reasons.
For more details as well as a computation
of $\zeta_D(0)$ and $\zeta\p_D(0)$ for DBC in a thermal bath,
see \cite{Luiz}.

\bigskip

LCA would like to thank Adilson J. da Silva, Marcelo Gomes and
 Carlos Farina for reading the manuscript,
and also to the Mathematical Physics Department  for their kind 
hospitality.  This work was supported by  the foundation
FAPESP.

\end{document}